\documentclass{svproc}
\usepackage{url}
\usepackage{graphicx}

\begin{document}
\mainmatter              
\title{Hadronic resonances production with ALICE at the LHC}
\titlerunning{Hadronic resonances production with ALICE at the LHC}  
%
\author{Sushanta Tripathy (for the ALICE collaboration)}
\authorrunning{Sushanta Tripathy (for the ALICE collaboration)} 
%
\tocauthor{Sushanta Tripathy (for the ALICE collaboration)}
\institute{Discipline of Physics, School of Basic Sciences, Indian Institute of Technology Indore, Simrol, Indore-453552, India\\
\email{Email: Sushanta.Tripathy@cern.ch}}
\maketitle              
\begin{abstract}
Measurements of the production of short-lived hadronic resonances are used to probe the properties of the late hadronic phase in ultra-relativistic heavy-ion collisions. Since these resonances have lifetimes comparable to that of the fireball, they are sensitive to the competing effects of particle re-scattering and regeneration in the hadronic gas, which modify the observed particle momentum distributions and yields after hadronisation. Having different masses, quantum numbers and quark content, hadronic resonances carry a wealth of information on different aspects of ion-ion collisions, including the processes that determine the shapes of particle momentum spectra, insight into strangeness production and collective effects in small collision systems. We present the most recent ALICE results on $\rho(770)^{0}$, K*(892)$^{0}$, $\phi(1020)$, $\Sigma(1385)^{\pm}$, $\Lambda(1520)$, $\Xi(1530)^{0}$ and $\Xi(1820)$ production at the LHC. They include measurements performed in pp, p--Pb and Pb--Pb collisions at different energies, as well as the latest results from the LHC Run 2 with Xe--Xe collisions at $\sqrt{s_{\rm NN}}$ = 5.44 TeV and with Pb--Pb collisions at $\sqrt{s_{\rm NN}}$ = 5.02 TeV. Collision energy, centrality and multiplicity differential measurements integrated yields and particle ratios are discussed in detail. A critical overview of these results are given through comparisons to measurements from other experiments and theoretical models. 

\keywords{Hadronic resonances, Strangeness production, Re-scattering}
\end{abstract}
\section{Introduction}
Hadronic resonances are very interesting probes to study the properties of the hadronic medium formed in ultra-relativistic heavy-ion collisions, as the yield ratios of resonances to stable hadrons provide information about the re-scattering and regeneration effects in the hadronic medium~\cite{Bleicher:2002dm,Bleicher:2003ij}. Resonances with shorter lifetime are expected to decay inside the hadronic phase formed in ultra-relativistic heavy-ion collisions and if the elastic or pseudo-elastic scattering of their decay products (re-scattering) is dominant over regeneration, the resonance yield after kinetic freeze-out would be smaller than the one originally produced at the chemical freeze-out. The re-scattering and regeneration processes may also cancel each other. If the lifetimes of resonances are more than the lifetime of the hadronic phase, their yields will not be affected by any such processes. Along with the insight into the hadronic phase of the system, resonances with open and hidden strangeness shed light on strangeness production in different collision systems. It is expected that the particles with open strangeness may be subject to canonical suppression in small collision systems with respect to large systems. However the $\phi$ meson, a hidden-strangeness particle, is not
expected to be canonically suppressed~\cite{Vislavicius:2016rwi}.

We present the most recent ALICE results on $\rho(770)^{0}$, K*(892)$^{0}$, $\phi(1020)$, $\Sigma(1385)^{\pm}$, $\Lambda(1520)$, $\Xi(1530)^{0}$ and $\Xi(1820)$ production at the LHC. They include measurements performed in pp, p--Pb and Pb--Pb collisions at different energies, as well as the latest results from the LHC Run 2 with Xe--Xe collisions at $\sqrt{s_{\rm NN}}$ = 5.44 TeV and with Pb--Pb collisions at $\sqrt{s_{\rm NN}}$ = 5.02 TeV.

\begin{figure}[ht!]
\centering
\includegraphics[height=28em]{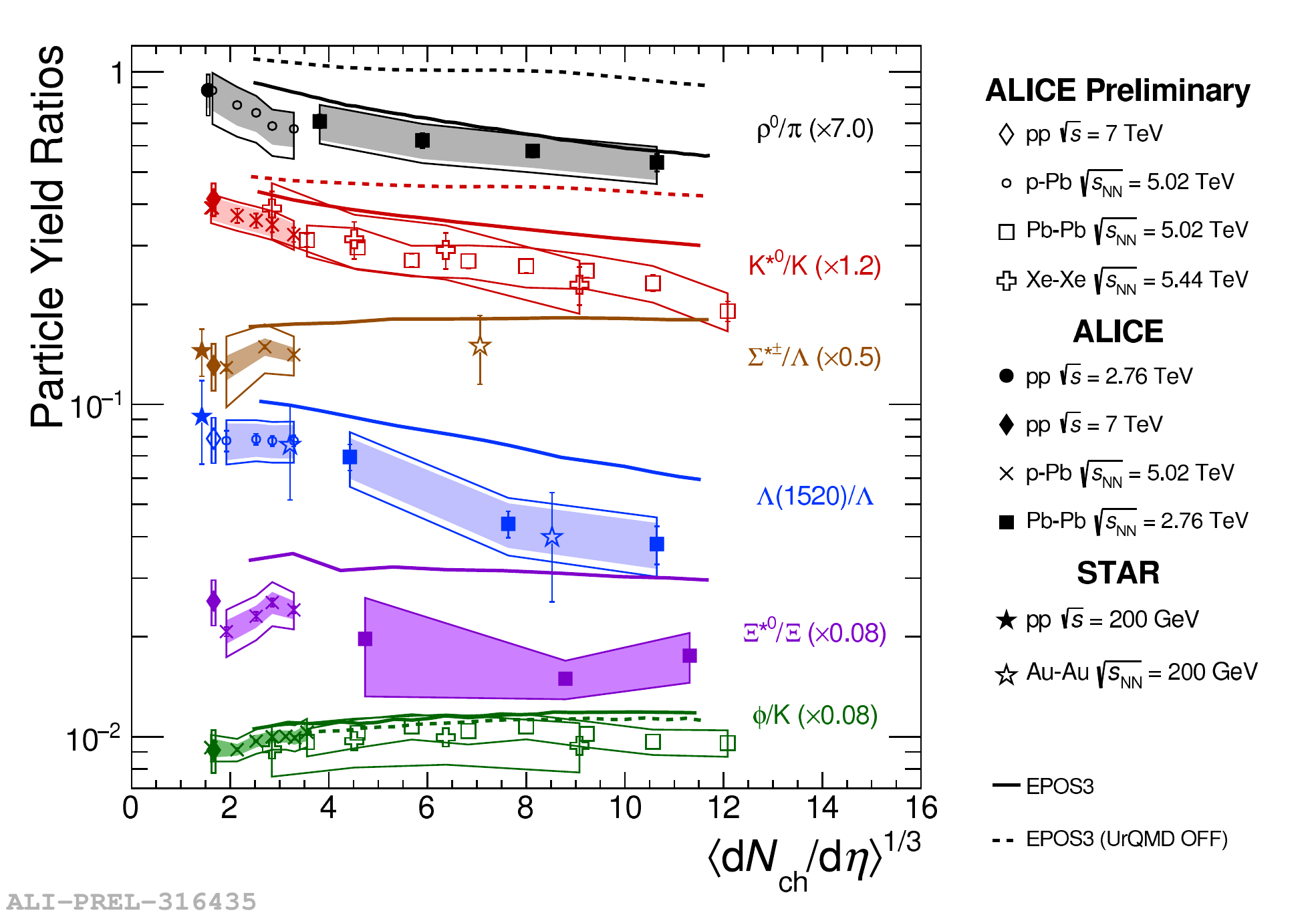}
\caption{Summary of particle yield ratios of different resonances to their respective ground-state particles as a function of multiplicity for pp, p--Pb, Xe--Xe and Pb--Pb collisions along with comparisons to EPOS3 predictions and STAR data.}
\label{fig1}
\end{figure}

\section{Results and discussions}

Figure~\ref{fig1} shows the particle yield ratios of different resonances to their ground states (ordered by increasing resonance lifetime from top to bottom) as a function of multiplicity for pp, p--Pb, Xe--Xe and Pb--Pb collisions. Comparisons with EPOS3~\cite{Knospe:2015nva} and STAR data are also shown in Fig.~\ref{fig1}. A significant suppression of the $\rm{\rho/\pi}$, $\rm{K}^{*0}/K$ and $\rm{\Lambda^{*}/\Lambda}$ ratios has been observed with increasing charged particle multiplicity. This indicates the dominance of re-scattering over regeneration for $\rm{\rho}$, $\rm{K}^{*0}$ and $\rm{\Lambda}$. However, the long-lived resonance ratios like $\rm{\Xi^{*}/\Xi}$ and $\phi/\rm{K}$ are nearly constant as a function of multiplicity. This suggests the long-lived resonances are not significantly affected by the re-scattering or regeneration processes. It indicates that the long lived resonances decay predominantly outside the hadronic medium. The EPOS3 calculations without UrQMD seem to describe only the $\phi/\rm{K}$ while they fail to explain $\rm{\rho/\pi}$ and $\rm{K}^{*0}/K$ ratios. However, we observe that EPOS3 with UrQMD, which includes a modelling of re-scattering and regeneration in the hadronic phase, seems to explain all the particle ratios qualitatively.

\begin{figure}[ht!]
\centering
\includegraphics[height=17em]{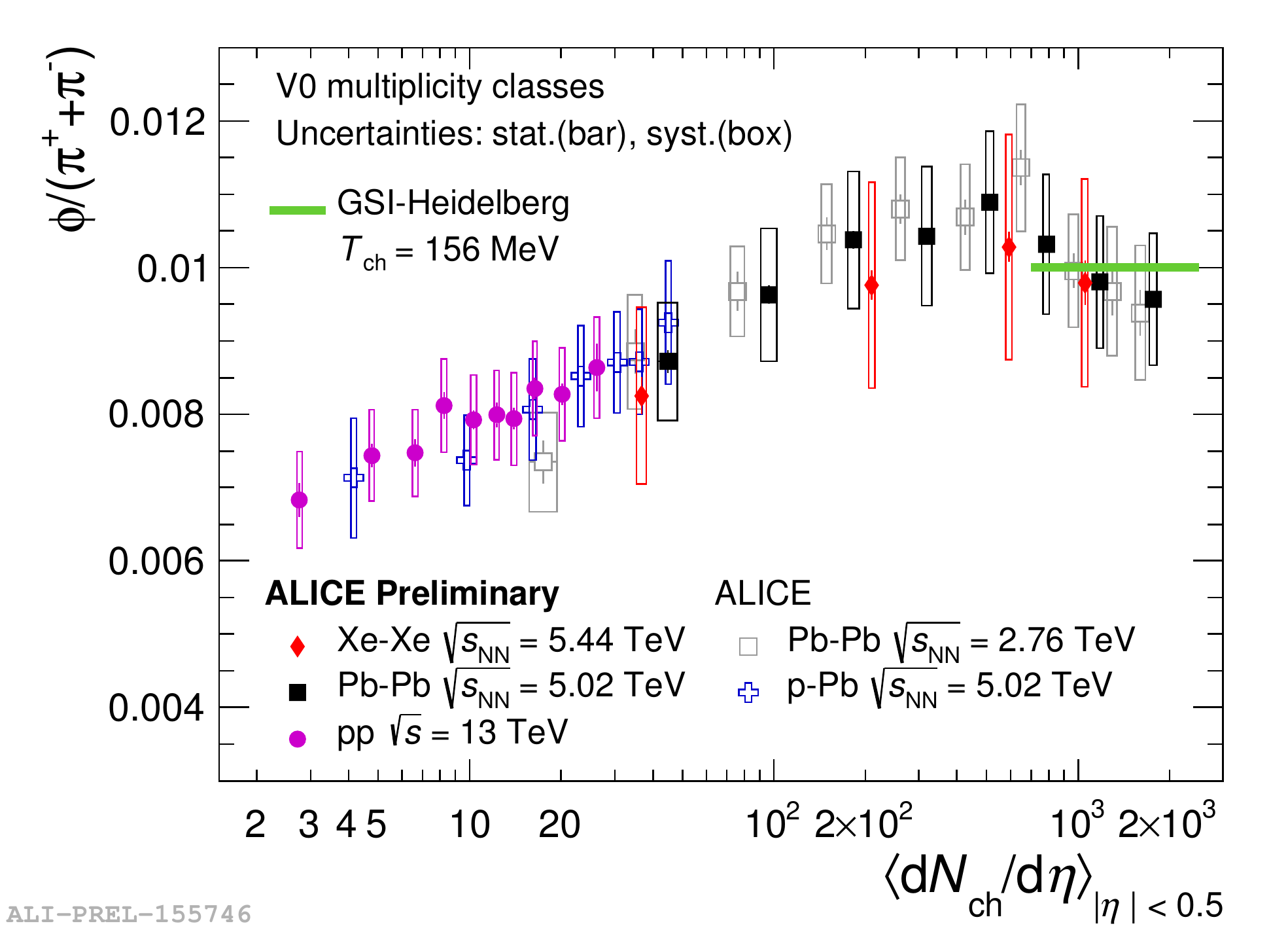}
\caption{$p_{\rm T}$-integrated $\phi$/($\pi^{+}+\pi^{-}$) ratio in pp, p--Pb and Pb--Pb collisions as a function of charged-particle multiplicity~\cite{Abelev:2014uua,Adam:2016bpr}. Statistical uncertainties are represented by bars and total systematic uncertainties by open boxes. The Grand Canonical thermal model prediction~\cite{Andronic:2017pug} is shown as solid green line.}
\label{fig2}
\end{figure}

\begin{figure}[ht!]
\centering
\includegraphics[height=17em]{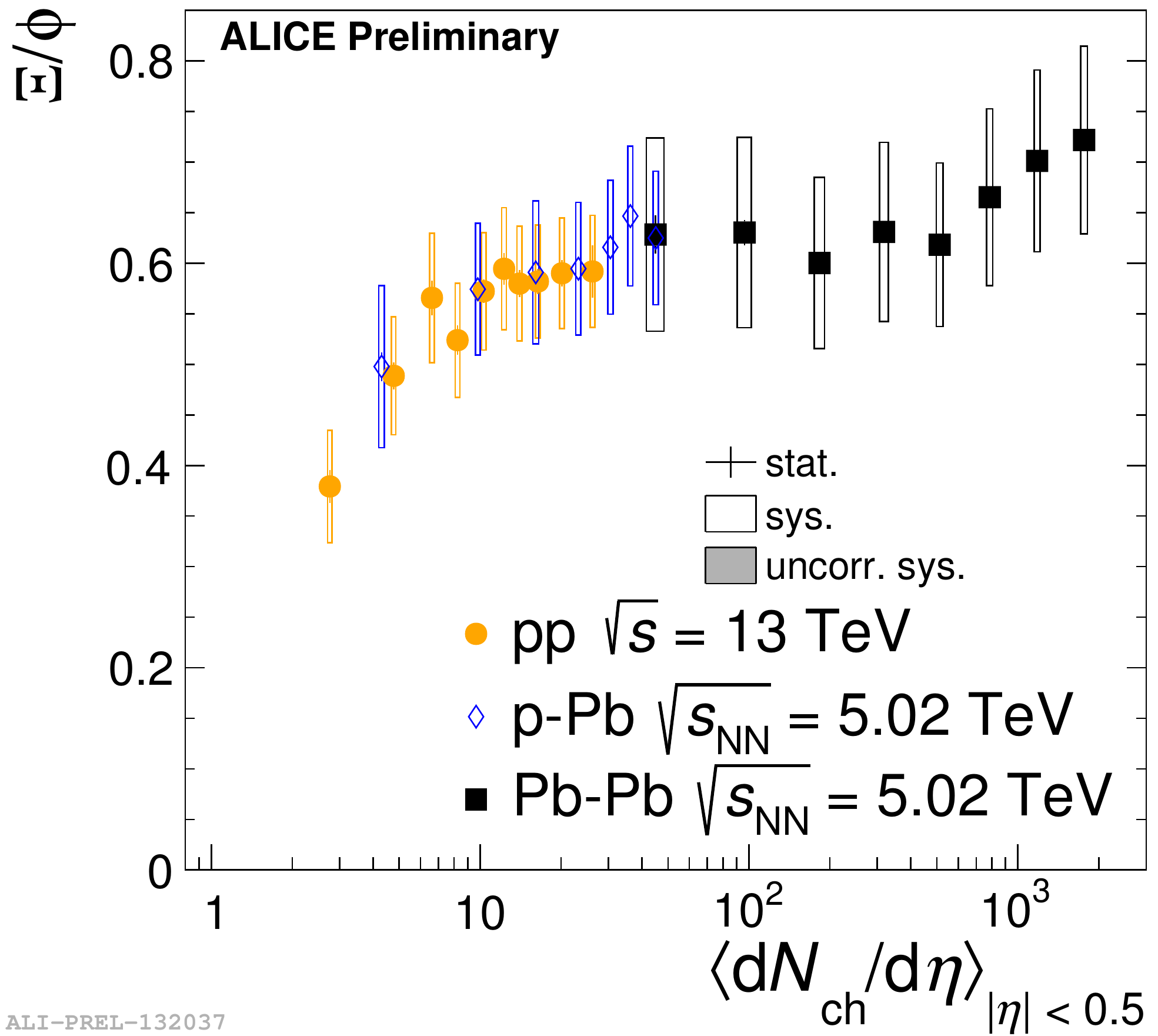}
\caption{$p_{\rm T}$-integrated $\rm{\Xi}/\phi$ ratio in pp, p--Pb and Pb--Pb collisions as a function of charged-particle multiplicity. Statistical uncertainties are represented by bars and total systematic uncertainties by open boxes.}
\label{fig3}
\end{figure}

Figures~\ref{fig2} and~\ref{fig3} show the $p_{\rm T}$-integrated $\phi$/($\pi^{+}+\pi^{-}$) and $\rm{\Xi}/\phi$ ratios, respectively in pp, p--Pb and Pb--Pb collisions as a function of charged-particle multiplicity. These ratios give insight into strangeness production using the $\phi$ meson, a hidden-strangeness particle. The ratio $\phi$/$\pi$ increases as a function of multiplicity in small collision systems and approaches the thermally predicted value~\cite{Andronic:2017pug} at high multiplicity in large collision systems. Even though the $\phi$ meson has hidden strangeness, this behavior is consistent, against the expectations with its canonical suppression in small systems~\cite{Vislavicius:2016rwi} and it favors the non-equilibrium production of $\phi$ and/or strange particles. The $\phi/\rm{K}$ (Fig.~\ref{fig1}) and $\rm{\Xi}/\phi$ ratios remain fairly flat across a wide multiplicity range. These ratios suggest that the $\phi$ behaves similarly to particles with open strangeness.

\section{Summary}
In summary, the ratios of $\rm{\rho/\pi}$, $\rm{K}^{*0}/K$ and $\rm{\Lambda^{*}/\Lambda}$ show a decreasing trend from pp and peripheral Pb--Pb to central Pb--Pb collisions. This suggests the dominance of re-scattering over regeneration processes for short-lived resonances in the hadronic phase of the system formed in heavy-ion collisions. However, yield ratios for long-lived resonances remain flat as a function of multiplicity. This suggests that long-lived resonances are not significantly affected by re-scattering or regeneration processes, or that those processes may cancel each other out. It indicates that the long lived resonances decay predominantly outside the hadronic medium. EPOS3 with UrQMD seems to reproduce these trends qualitatively. The $\phi$ meson, a hidden-strangess particle, shows similar behavior to particles with open strangeness and seems to have an effective strangeness between 1 and 2~\cite{Tripathy:2018ehz,Song:2018rdl}.
%
%

\end{document}